\documentclass[conference]{IEEEtran}
\IEEEoverridecommandlockouts

\usepackage{cite}
\usepackage{amsmath,amssymb,amsfonts}
\usepackage{algorithmic}
\usepackage{graphicx}
\usepackage{textcomp}
\usepackage{xcolor}
\def\BibTeX{{\rm B\kern-.05em{\sc i\kern-.025em b}\kern-.08em
    T\kern-.1667em\lower.7ex\hbox{E}\kern-.125emX}}
\begin{document}

\title{Data-driven Modeling for Grid Edge IBRs: A Digital Twin Perspective of User-Defined Models \\
}

\author{\IEEEauthorblockN{Kaveri Mahapatra*, Bhaskar Mitra, and Soumya Kundu}
\IEEEauthorblockA{\textit{Pacific Northwest National Laboratory, United States} \\
\{kaveri.mahapatra, bhaskar.mitra, soumya.kundu\}@pnnl.gov}
% \and
% \and
% \and
% \and
% \and
}

\maketitle

\begin{abstract}

Recent Odessa disturbance events \cite{nerc2021odessa, nerc2022odessa} have brought attention to the challenges associated with the interaction between Inverter-Based Resources (IBRs) and the transmission and distribution system. The NERC event diagnosis report has highlighted several issues, emphasizing the need for continuous performance monitoring of these IBRs by system operators. Key areas of concern include the mismatch of control and protection performance of IBRs between the original equipment manufacturer (OEM)-provided models and field measurements. The inability to replicate the realistic response can result in incorrect reliability and resilience studies. In this paper, we developed an approach on how to emulate the behavior of an IBR using measurement data obtained for system operators to utilize in real-time and long-term planning. Two experiments are conducted in the phasor domain and electromagnetic transients (EMT) domain to emulate the behavior for grid forming and grid following inverters under various operating conditions and the effectiveness of the proposed model is demonstrated in terms of accuracy and ease of utilizing user-defined models (UDMs).
\end{abstract}


\section{Introduction}

Precise modeling of inverter-based resources (IBRs)  and distributed generation is necessary for energy systems planning studies and real-time operations. Model accuracy is essential to prevent reliability concerns with interconnections and streamline the interconnection process. Aligning the models of grid-edge assets with high fidelity from measurements ensures accurate control and protection representation and enhances model usability in various planning studies.

 % maintaining the reliability of bulk energy systems (BES).

% Digital twins have the flexibility to employ various models that accurately represent the physical objects they replicate. In an optimal scenario characterized by online computation and high precision, digital twins would utilize models directly derived from physics. These models would comprehensively incorporate all relevant phenomena influencing the measurement and seamlessly update as needed through continual model validation and model calibration. For example, the digital twin of an IBR would be able to simulate the control performance under various operating conditions in real-time and update the knowledge about control modes based on real-time measurements so real-time situational awareness can be improved over the IBR operations. However, in the absence of physics information from OEMs, only black box models can be prepared from measurements. % Reduce

Absence of the proprietary original equipment manufacture (OEM) information \cite{nerc2020BPS} may require transmission operators (TOPs) to calibrate the IBR generic models \cite{esig_generic}. This process requires solving a nonlinear optimization problem with extensive event filtering, event selection, sensitivity analysis, parameter selection, and manual adjustments of parameters. Generic models calibrated based on single events are often inadequate \cite{km_gen_model,ju2020indices} to represent a large set of event responses of IBRs. This problem requires continual model validation and calibration methods of the IBR and an understanding of the current behavior before applying control actions to the system. Preparing a 'heavily parametric' IBR model as suggested in recent FERC order 901 \cite{ferc_901} could be very computationally expensive for generic models with insufficient physics and thus may be only suitable for post-event data performance assessment. It is not "realistic" for fast time scale real-time decision-making with changing operational modes. In this direction, data-driven surrogate/metamodels \cite{kim2020machine} or digital twin models \cite{dig_twin} could serve as an excellent platform for representing the online behavior of different types of IBRs based on online data ingestion. An additional benefit arises when the surrogate model is assessed at input values where the outcome of the full model is uncertain. It not only provides an estimation of the model's result for those values but also offers an estimation of the associated error and model performance. The error estimate helps pinpoint areas within the input space where obtaining the model result would be most advantageous in minimizing error. This enables iterative refinement of the surrogate model, concentrating efforts on areas where reducing error offers the most substantial benefits.

Therefore, we propose a method for online model estimation of IBR that leverages input-output response patterns obtained from measurements of IBRs under different modes of operation. The proposed approach aims to develop measurement-driven online digital twin/ user-defined models (UDMs) capable of closely capturing a broad spectrum of IBR dynamical information and modeling the uncertainty in the prediction. The calibration problem is framed as a nonlinear system identification problem. This work utilizes phasor data and point-on-wave data accessible from IBR or OEM-supplied model terminals often without access to control or reference parameters. The development of such models relies solely on continuous recordings obtained from IBR sites thus eliminating the need of controlled tests of IBRs. These models are expected for efficient continual model validation, calibration, online decision-making, and virtualization. this in turn enhances situational awareness of IBRs connected to the grid. 

The paper is organized as follows. Section II discusses problem formulation for IBR model calibration. Section III presents the proposed approach along with evaluation criteria. Results with grid-forming inverter (GFM), and grid following inverter (GFL) with both phasor-based and electromagnetic transients (EMT) domain models are presented in Section IV. Section V concludes the paper.

% \begin{itemize}
% \item Plant control updates
% \item IBR model updates
% \item PLL loss of synchronism
% \item Protection events
% \end{itemize}

\section{Problem Formulation} %(0.5 page)

%A standard representation of a power system involves nonlinear differential and algebraic equations that capture the dynamic interplay among various components and generators. 

%To represent a power plant, the set of measurements that represent plant response to various grid events must be made available. 

Modeling an IBR from measurements requires the following steps. (1) Extract the set of measurements necessary for the IBR model calibration and validation such as input and output variable measurement streams; (2) Identify a simulation environment for building the IBR model; (3) Prepare IBR model codes playback in the simulation environment to pass on the input streams and predict the output streams; (4) Calibrate and validate the UDM codes using those streams with a set of prediction error metrics; (5) Evaluate the prediction error from the UDM of IBRs. To prepare a UDM in a simulation environment aimed at emulating the IBR model response, the variables linked to the IBR model from those about the network need to be segregated. One common approach to accomplish this task is by recording the algebraic variables like voltage ($V$), frequency ($f$), as well as active and reactive power ($P$, $Q$) at the points where the IBR connects to the network.

For positive sequence modeling, in the case of grid following (GFL) IBR configurations, the calibration platform must be structured to enable the playback of recorded variables such as voltage ($v$) and frequency ($f$) obtained from the network, alongside simulating the IBR model described by parameters ($\bf{p}$). This setup allows for the estimation of active ($P$) and reactive ($Q$) powers at the terminals of the IBR. When the estimated parameters closely approximate the true parameters of the IBR model used during data generation, the estimated powers ($\hat P$ and $\hat Q$) should correspond closely to the recorded values of $P$ and $Q$. In the absence of system connections to the IBR, and model structure, control modes, and references, the calibration problem can be formulated as similar to that of a black-box model identification which does not require the availability of any physics information to estimate dynamic surrogate model.

Similarly, in the case of a grid-forming IBR configuration, the calibration platform must be structured to enable the playback of recorded variables such as active ($P$) and reactive ($Q$) powers at the terminals of the IBR, alongside simulating the IBR model described by parameters ($p$). This setup allows for the estimation of voltage ($V$) and frequency ($f$) obtained from the network. %When a set of parameters closely approximates the true parameters of the IBR model used during data generation, the estimated voltage and frequency ($\hat V$ and $\hat f$) should correspond closely to the recorded values of $V$ and $f$. 

\begin{figure}
    \centering
    \includegraphics[clip,width=60mm,scale=0.2]{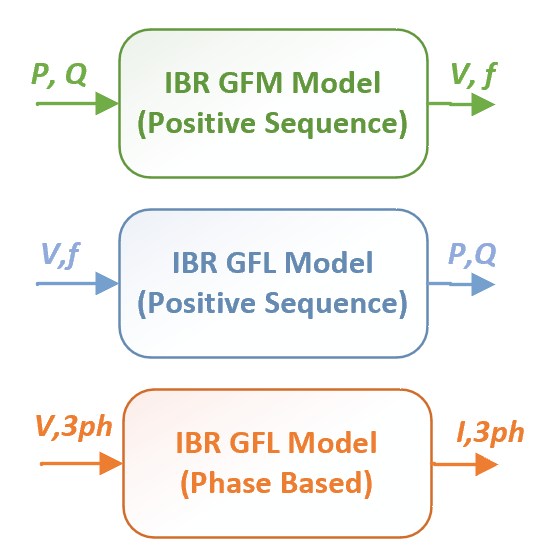}
    \caption{IBR playback modeling- A User-Defined Model in python}
    \label{fig:UDM_fig}
\end{figure}

\begin{figure}
    \centering
    \includegraphics[clip,width=80mm,scale=0.4]{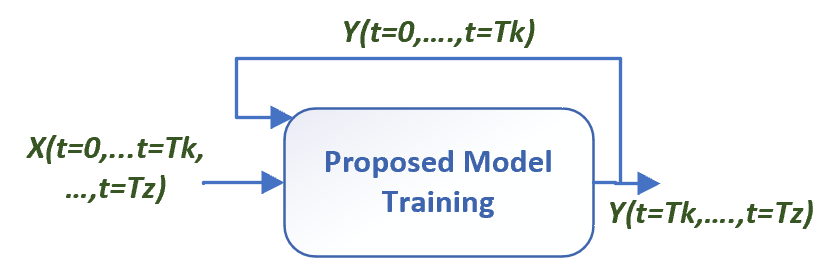}
    \caption{Multi-step time series recursive prediction modeling for IBRs}
    \label{fig:multistep}
\end{figure}

\begin{figure*}[t]
    \centering
    \includegraphics[clip,width=180mm,scale=0.8]{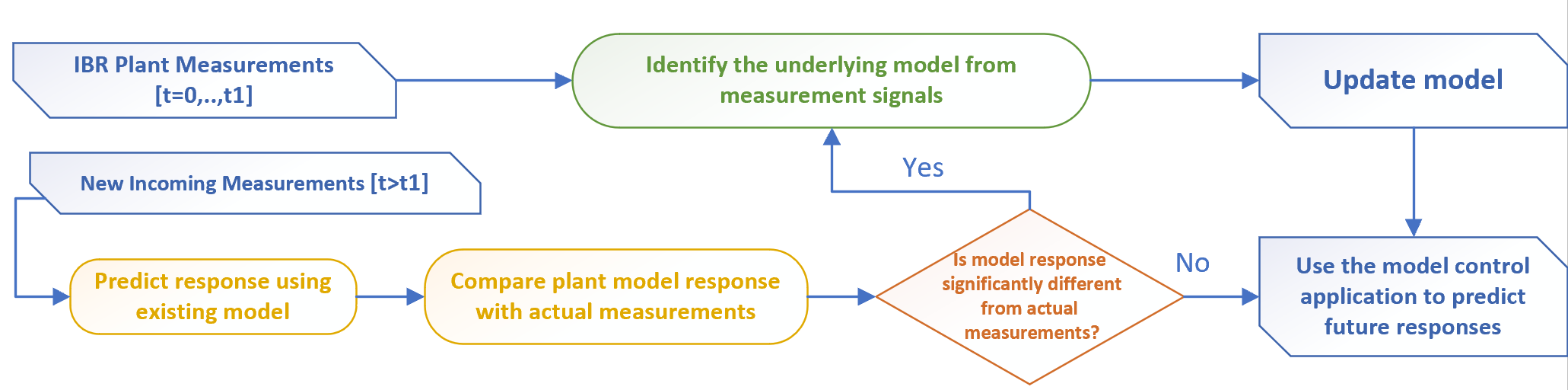}
    \caption{Flowchart describes the step-by-step procedure to translate IBR measurements into a UDM using the proposed approach}
    \label{fig:flow_chart_v1}
\end{figure*}

For phase-based modeling in EMT domain simulation of IBRs, the calibration platform must be structured to enable the playback of recorded variables such as voltage 3 phase recordings ($v_{abc}$) and estimate the currents ($i_{abc}$) injected by IBRs \cite{fan2020time}. Figure \ref{fig:UDM_fig} shows the input and output attributes for each type of IBR modeling. Therefore, given a set of input and output recorded measurements from an IBR under various operating conditions and disturbances, the IBR model identification problem can be formulated as a system identification-based optimization problem as shown in equation \ref{eq:obj}. The solution gives a set of parameters for UDM which can be used for predicting IBR's transient response to a disturbance event occurring in a grid.

\begin{equation}
\label{eq:obj}
\mathop {\min }\limits_p \sum\limits_{m = 1}^{{N_m}} {\sum\limits_{t = {t_1}}^{{N_T}} {\left( {{{{{\hat y}}}_{m,t}} - {{{y}}_{m,t}}} \right)} }
\end{equation}

\begin{equation}
{{{{\hat y}}}_{m,t}} = {\mathop{\rm f}\nolimits} \left( {{\bf{u}},{{\bf{y}}_{t - 1:{N_T}}},{\bf{p}}} \right)
\end{equation}

% min sum ((f(p[0,t])-Y[0,t])) with p. 

% where f is the outout of a nonlinear function. 

% Where, $f$ is formulated as an ARMAX model in this problem, and $\bf{p}$ is the coefficients of the functions. %A network has been used for this problem. 
The function $\bf{f}$ is defined as a nonlinear dynamical system of equations, where $p$ represents the coefficients/parameters of the model described by function $\bf{f}$. Here, $N_m$ and $N_T$ denote the number of output and input time series signals of the plant, respectively. At a given time $t$, $u_{(m,t)}$ indicates the $m^{th}$ input signal to the plant. The variables $y_{(m,t)}$ and $\hat y_{(m,t)}$, respectively, denote the actual response obtained from measurements and the predicted response from the estimated model $\bf{f}$.

The parameters $\bf{p}$ can be obtained generally by solving the nonlinear optimization problem using a stochastic gradient descent algorithm, which aims to minimize the difference between the actual and predicted responses of the UDM, as expressed by a loss function. To address this multistep prediction problem, we propose using an autoregressive (AR) structure for the model, which can serve as a UDM for IBRs.

% \begin{figure}
%     \centering
%     \includegraphics[clip,width=90mm,scale=0.5]{Figures/response_pred1.PNG}
%     \caption{Caption}
%     \label{fig:enter-label}
% \end{figure}

\section{Proposed Approach} %(1-1.5 page)
In this work, we proposed an Auto-Regressive Moving Average with eXogenous inputs (ARMAX) model \cite{yang1995identification} for the continuous representation of IBR responses. ARMAX method for continual model validation and model updates is proposed in this work which is a type of recursive time series model structure used for forecasting and analysis. It can describe the relationship between a set of time series variables (the response) and one or more input series (exogenous variables), along with its own past response values and past errors as shown in Figure \ref{fig:multistep}. An ARMAX model incorporates autoregressive (AR) and moving average (MA) components along with exogenous (X) variables to capture the dynamics of a time series and make predictions based on both the historical and incoming data and external factors. An ARMAX model can be represented as follows \cite{yang1995identification}.

\begin{equation}
{y_t} = \sum\limits_{i = 1}^{{N_p}} {{\alpha _i}{y_{t - i}}}  + \sum\limits_{i = 1}^{{N_q}} {{\beta _i}{\varepsilon _{t - i}}}  + \sum\limits_{i = 1,j = 1}^{{N_q},{N_m}} {{\gamma _i}{x_{t,j}}}  + {e_i}
\end{equation}

where, $y_t$, $e_t$, $x_t$ is the dependent variable, error term, and exogenous variable at time $t$. ${\alpha _i}$ is the autoregessive parameter,  ${\beta _i}$ is the moving average parameter, ${\gamma _i}$ is the exogenous weight parameter. $N_p$ is the order of the autoregressive part, $N_q$ is the order of the moving average part. $N_m$ is the number of exogenous variables.

% The ARMAX model is particularly useful for modeling time series data where the relationship between variables is influenced by both past values of the dependent variable and exogenous factors. It is commonly employed in various fields such as economics, engineering, and signal processing for tasks such as forecasting, system identification, and control.

Note that ARMA models have been used in power system modal estimation \cite{wecc_arma} which involves extracting different inter-area mode frequencies from phasor data by estimating an Auto-regressive Moving Average (ARMA) models. Inspired by those concepts, for representing the behavior of IBRs, which also requires phasor data from a network, an ARMAX model structure is proposed in this work. For building a UDM for IBRs with continual validation using ARMAX, a set of steps are proposed as shown in flow chart in Figure \ref{fig:flow_chart_v1} and can be described as follows. Continuous model validation and calibration using IBR terminal measurements involve an integrated workflow starting with the real-time acquisition of data from plant sensors, which is then stored in a time-series database. This data is preprocessed through scaling and normalization to ensure consistency. An initial model is calibrated with historical data to establish a baseline identification model. This model continuously operates in real time, generating predictions which are compared with actual plant measurements to calculate errors. Performance metrics such as Mean Root Mean Squared Error (RMSE) is monitored continuously. When significant deviations are encountered in IBR-predicted responses, an automated recalibration of the ARMAX model is triggered. This feedback loop ensures the model's accuracy and stability, supported by real-time dashboards for monitoring and detailed logs for documentation. Periodic analysis and updates to the model can track the operational changes and can incorporate new changes in the system dynamics. It is expected that continual model validation for IBR models can maintain the identified model's integrity and reliability in real time. The following section discusses the experimental results obtained under continual model validation. 

\begin{figure}[thpb]
    \centering
    \includegraphics[scale=0.6]{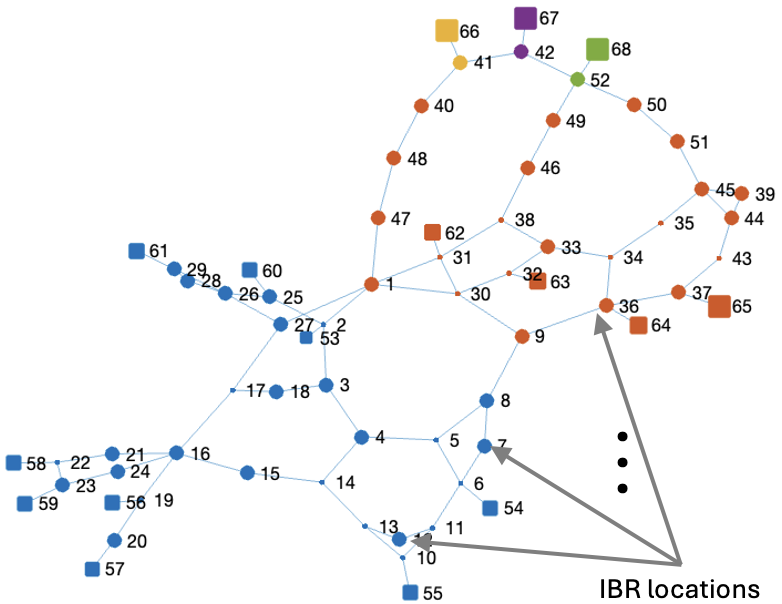}
    % \vspace{-30pt}
    \caption{Schematic of the phasor-domain positive sequence test model (modified IEEE 68-bus network) with 35 IBRs co-located at the load buses, marked by circles. The network has 16 generators, marked by squares, across 5 different regions (marked by different colors).}
    % \vspace{-20pt}
    \label{fig:network_IBR}
\end{figure}

\section{Results} %(1.5-2 pages)

% \subsection{Test system description for positive sequence model}

% {\color{red}system description, how the data was generated, what events were simulated and automated?

% Soumya - system description and data generation}

This section discusses on the results obtained from testing the proposed approach on different test systems and for different types of IBRs. It is divided thus into two sections, (1) phasor domain modeling; (2) EMT domain modeling. The description of the test system along with the data collection and processing steps used for testing the proposed application is described in the following sections.

\subsection{Phasor domain modeling of IBR}

\subsubsection{Test system description}

One of the test systems used to numerically evaluate the performance of the IBR modeling algorithm is a phasor domain (positive sequence) model involving multiple IBRs connected over a power transmission network. Specifically, in this work, a modified IEEE 68-bus test network is used, shown in Fig.\,\ref{fig:network_IBR}, involving 35 IBRs connected to the network at the load buses \cite{singh2013report}. Each inverter has the same capacity rating, with their total (cumulative) capacity at 20\% of the total load on the network.

\begin{figure}
\vspace{-25pt}
    \centering
    \includegraphics[clip,width=80mm,scale=0.6]    {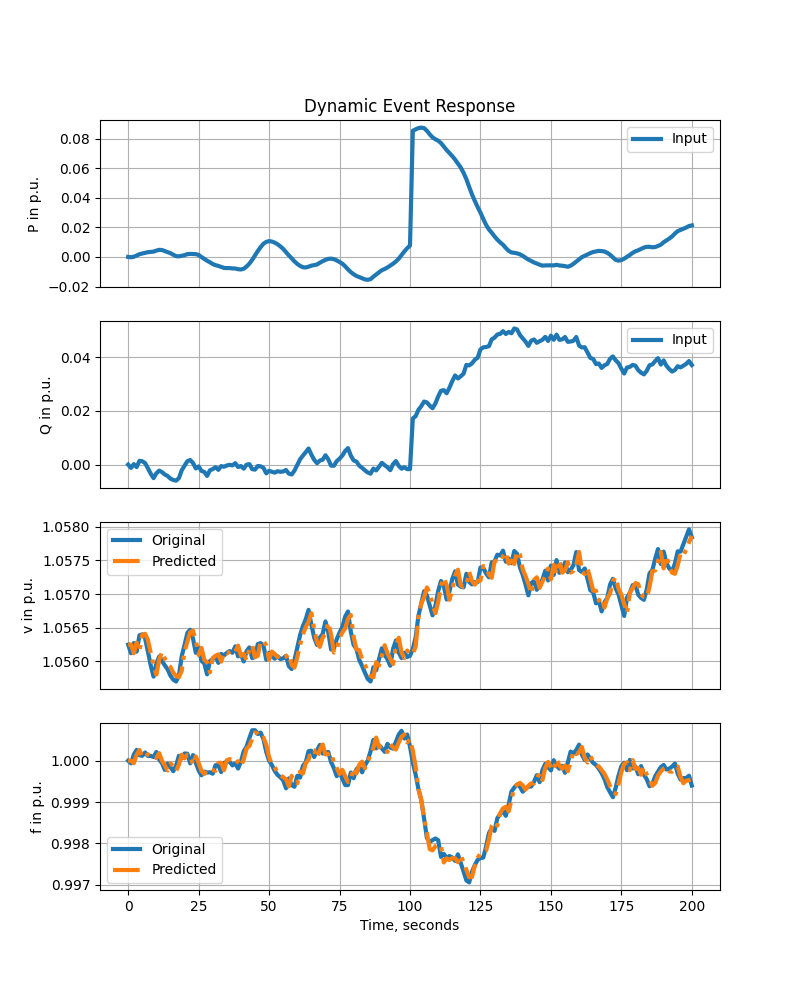}
    \vspace{-20pt}
    \caption{GFM IBR-1 response to a dynamic event in the network; Measurements are obtained from point of connection of IBR-1 with the network under a line trip event (event-A). The phasor domain response is in 1 ms resolution.}
    \vspace{-15pt}
    \label{fig:GFM_response_IBR_1_event_A}
\end{figure}
\begin{figure}
% \vspace{-20pt}
    \centering
    \includegraphics[clip,width=80mm,scale=0.6]    {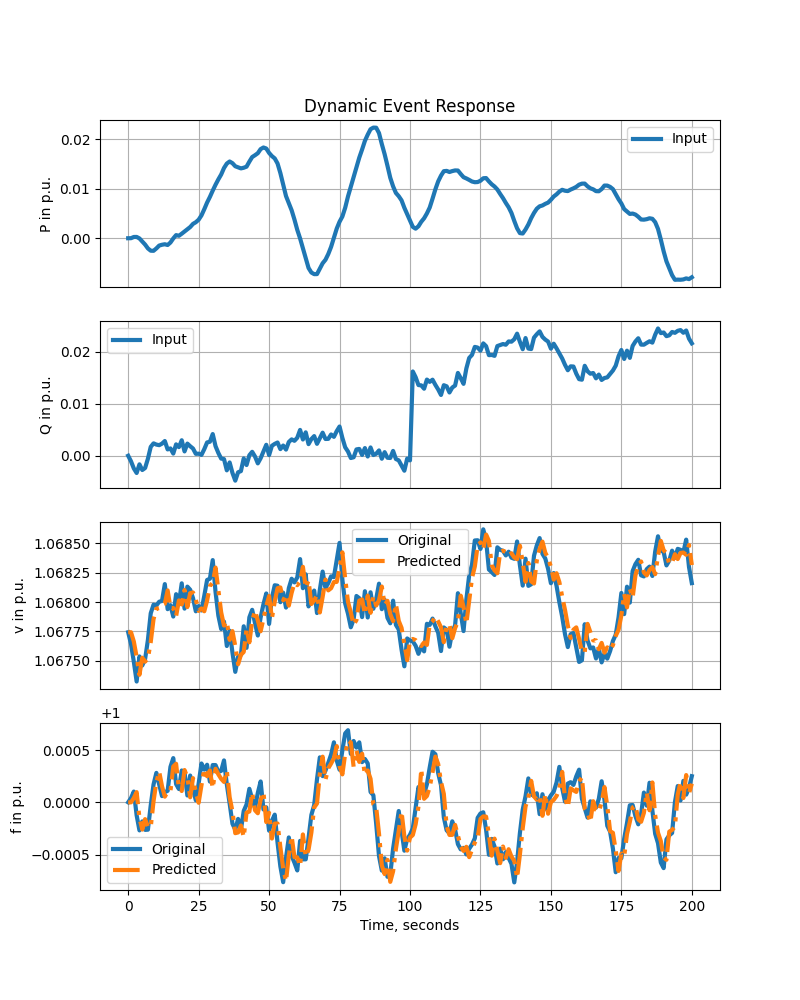}
    \vspace{-20pt}
    % {Figures/Ind_gfm_IBR_case_3_inv_20_r0.png}
    % {Figures/Ind_gfm_IBR_case_12_inv_21_r0.png}
    \caption{GFM IBR-2 response to a dynamic event in the network; Measurements are obtained from point of connection of IBR-2 with the network under a line trip event (event-B). The phasor domain response is in 1 ms resolution.}
    \vspace{-15pt}
    \label{fig:GFM_response_IBR_2_event_B}
\end{figure}
\begin{figure}
% \vspace{-15pt}
    \centering
    \includegraphics[clip,width=80mm,scale=0.8]{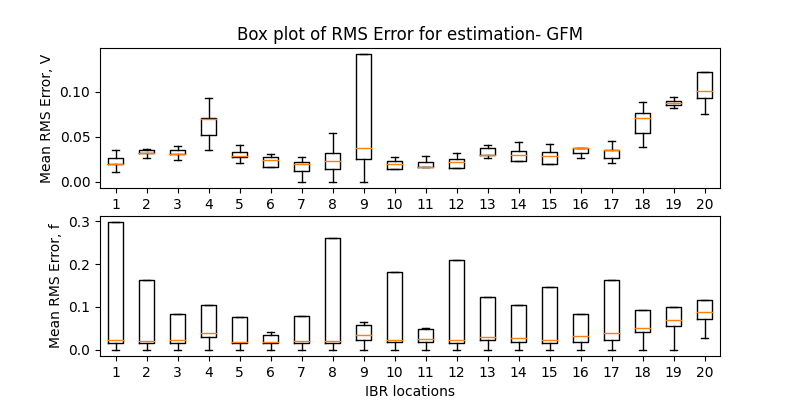}
    \vspace{-5pt}
    \caption{Distribution of mean RMS error in model validation across 25 events for 20 IBR (GFM) locations.}
    \vspace{-15pt}
    \label{fig:GFM_error_stat}
\end{figure}

Two specific scenarios are synthesized to evaluate the performance IBR modeling algorithm on the phasor-domain positive sequence network: 1) all IBRs modeled as grid-following inverters (GFLs), and 2) all IBRs modeled as grid-forming inverters (GFMs). The dynamic models of the GFL and GFM inverters are adopted from the existing literature \cite{lasseter2010certs,du2020modeling,kwon2023risk}. In particular, the GFLs are modeled as controllable current sources, with two key component blocks: a phase-locked-loop with a proportional-integral controller and a current control loop. On the other hand, the GFMs are modeled as per the CERTS droop control \cite{lasseter2010certs}, with an active power vs. frequency droop control loop and a reactive power vs. voltage droop control loop. Note protection techniques are not included as those were not part of \cite{lasseter2010certs,du2020modeling,kwon2023risk} modeling.

The datasets for learning and validating the ARMAX models of IBRs were generated by performing phasor-domain transient dynamic simulations of the network in response to different disturbance events, such as changes in load, line trips, and generation loss. A set of 25 contingency scenarios in the network are simulated in the phasor domain and responses from all the 35 IBRs are collected in each scenario. The test system was simulated in MATLAB and $V$, $f$, $P$, $Q$ measurements from the model were recorded. Each scenario data was normalized and scaled in the preprocessing steps to prepare those for ARMAX model training and testing. This work utilizes Systems Identification Package for PYthon (SIPPY) \cite{sippy_paper} platform to build ARMAX model for IBRs. This was used to build a simulation engine for IBR for input and output response feedback for recursive prediction of the model response. Both for GFM and GFL type of modeling, 2 input, 2 output, linear ARMAX state space model with output feedback was formulated using the ARMAX modeling structure. For identification of the system for IBRs, a recursive linear least squares solver was used to tune the parameters of the ARMAX model. The following section discusses the results obtained from phasor domain testing of the proposed model. Event scenarios with rich frequency content were selected for preparing the ARMAX model and the rest was used to test the prediction performance of the model.

\begin{figure}
    \centering
    \includegraphics[clip,width=80mm,scale=0.6]{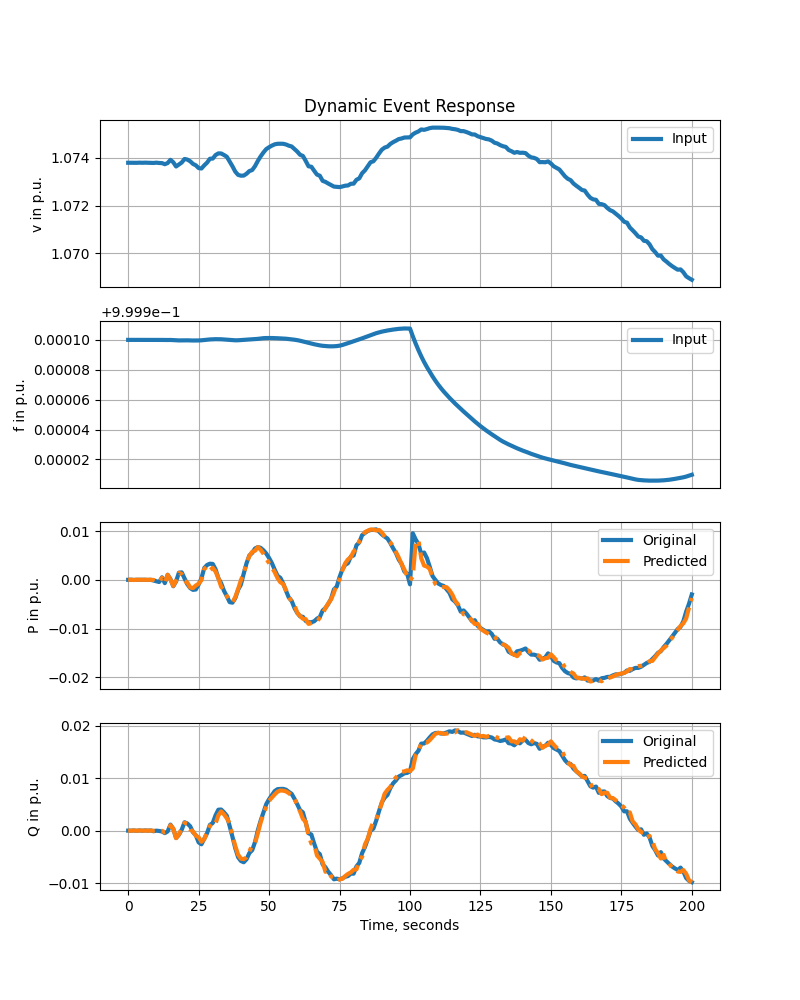}
    \vspace{-15pt}
    \caption{GFL IBR-1 response to a dynamic event in the network; Measurements are obtained from point of connection of IBR-1 with the network under a line trip event (event-C). The phasor domain response is in 1 ms resolution.}
    \vspace{-15pt}
    \label{fig:GFL_response_IBR_1_event_A}
\end{figure}

\begin{figure}
    \centering
    \includegraphics[clip,width=80mm,scale=0.6]{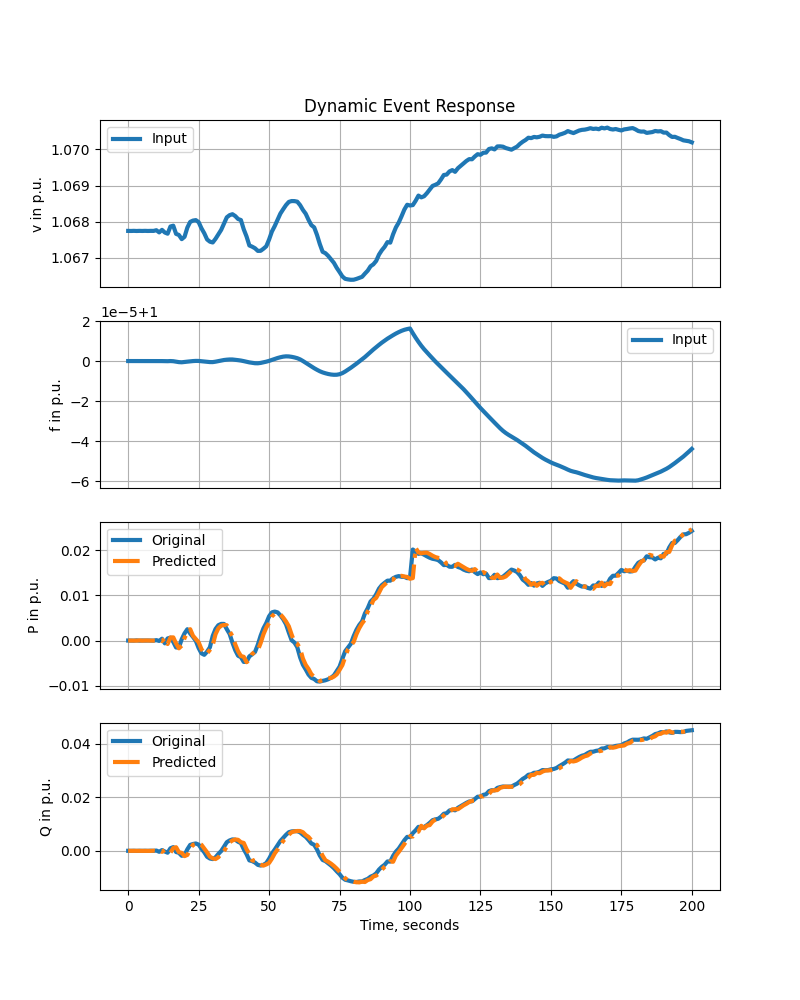}
    \vspace{-15pt}
    \caption{GFL IBR-2 response to a dynamic event in the network; Measurements are obtained from point of connection of IBR-2 with the network under a line trip event (event-D). The phasor domain response is in 1 ms resolution.}
    \vspace{-15pt}
    \label{fig:GFL_response_IBR_2_event_B}
\end{figure}

\begin{figure}
    \centering
    \includegraphics[clip,width=80mm,scale=0.8]{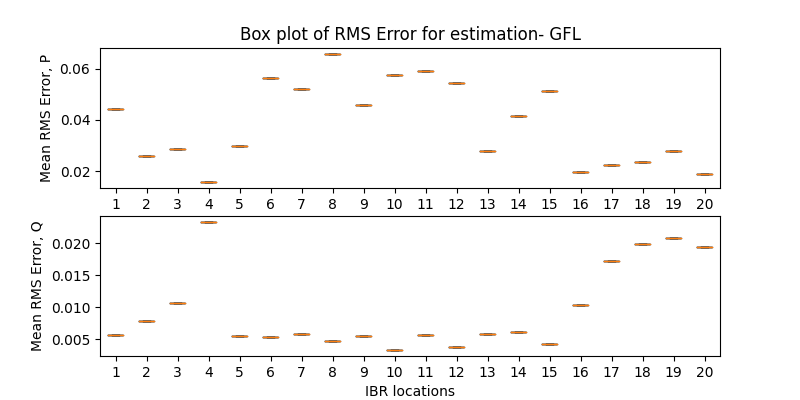}
    \caption{Distribution of mean RMS error in model validation across 25 events for 20 IBR (GFL) locations.}
    \label{fig:GFL_error_stat}
\end{figure}

\begin{figure}
    \centering
    \includegraphics[width=0.48\textwidth]{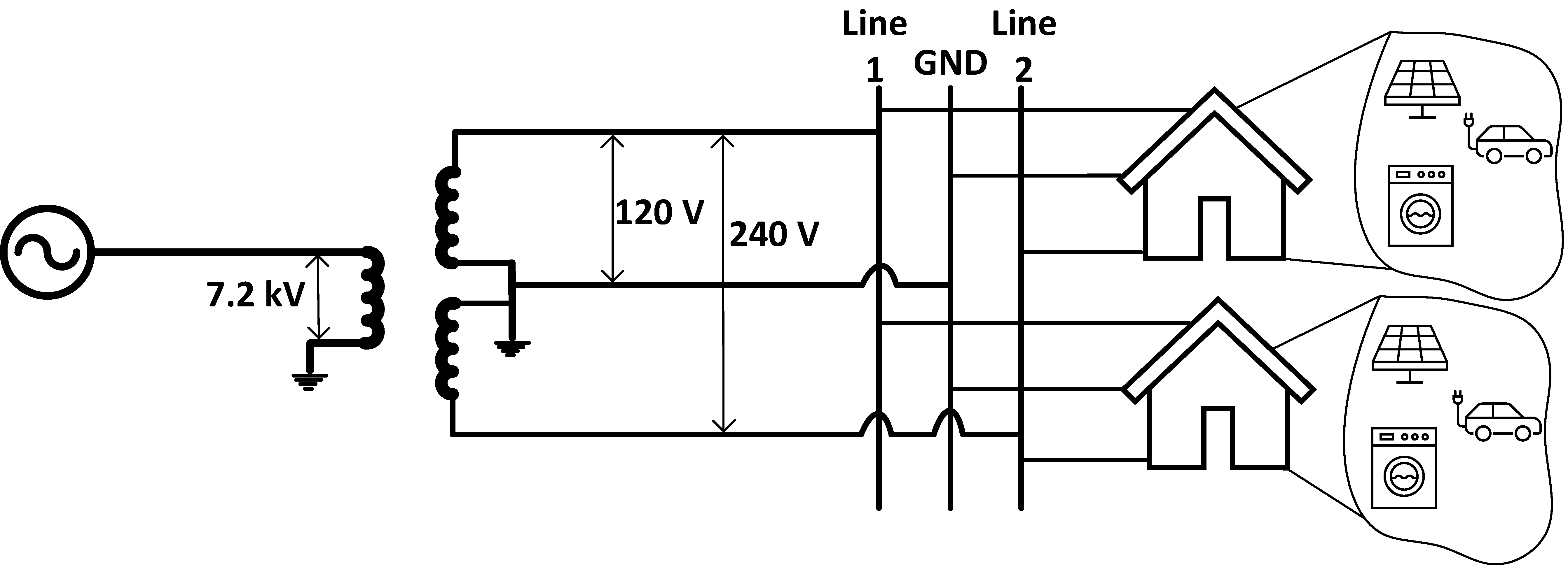}
    \caption{Schematic representing a setup for collection of voltage and current data for an EMT model.}
    \label{fig:EMT}
\end{figure}

\subsubsection{Phasor domain modeling results}
The results are shown for both GFM and GFL type of IBR model validation results. Fig. \ref{fig:GFM_response_IBR_1_event_A},\ref{fig:GFM_response_IBR_2_event_B} shows GFM modeling results with both input active and reactive power time series and output voltage and frequency response obtained under two line trip events in the test system. Both the true/original response from the test system and the estimated response from the ARMAX model for two different IBR locations are closely matching in this case. The RMS error estimated under 20 IBR models under 25 events during the model validation for GFM IBR ARMAX models are shown in Fig. \ref{fig:GFM_error_stat}. This figure shows the box plot of mean RMSE error plots for both $\hat V$ and $\hat f$. The distribution of RMSE errors shows lower average error rate across multiple validation responses and implies better accuracy of the ARMAX models GFM IBR phasor domain response estimation.

Fig. \ref{fig:GFM_response_IBR_1_event_A},\ref{fig:GFM_response_IBR_2_event_B} shows GFM modeling results with both input active and reactive power time series and output voltage and frequency response obtained under two line trip events in the test system. Both the true/original response from the test system and the estimated response from the ARMAX model for two different IBR locations are closely matching in this case. The RMS error estimated under 20 IBRs under 25 events during the model validation for GFM IBR ARMAX models are shown in Fig. \ref{fig:GFM_error_stat}. This figure shows the box plot of mean RMSE error plots for both $\hat V$ and $\hat f$. The distribution of RMSE errors shows lower average error rate across multiple validation responses and implies better accuracy of the ARMAX models GFM IBR phasor domain response estimation.

Figures \ref{fig:GFL_response_IBR_1_event_A} and \ref{fig:GFL_response_IBR_2_event_B} illustrate the results of GFL IBR ARMAX modeling, displaying both the input voltage and frequency time series, along with the output active and reactive power response during two line trip events in the test system. The actual responses from the test system closely match the estimated responses from the ARMAX model for two different IBR locations. Figure \ref{fig:GFL_error_stat} presents the RMS error estimated under 20 IBRs across 25 events during the validation of the identified GFM IBR models. This figure includes a box plot of the mean RMSE error plots for both \(\hat P\) and \(\hat Q\). The distribution of RMSE errors indicates a very low average error rate across multiple validation responses, suggesting better accuracy in the proposed ARMAX models' estimation of GFL IBR phasor domain responses.

% \begin{table}[]
%     \centering
%         \caption{Model Validation Error metrics in $\%$ for IBR in GFL configuration}
%     \begin{tabular}{c|c|c|c|c|c|c}
%     \hline
%     \hline
%         Response & IBR- & IBR- & IBR- & IBR- & IBR- & all \\
%         error & 1 & 2& 3 & 4 & 5 & \\
%         \hline
%         $e_{p_{\mu}}$& 1 & 2& 3 & 4 & 5 & all\\
%         \hline
%         $e_{q_{\sigma}}$& 1 & 2& 3 & 4 & 5 & all\\
%         \hline
%         $e_{p_{\mu}}$& 1 & 2& 3 & 4 & 5 & all\\
%         \hline
%         $e_{q_{\sigma}}$& 1 & 2& 3 & 4 & 5 & all\\
%         \hline
%     \end{tabular}
%     \label{tab:err_GFL_table}
% \end{table}

% \begin{table}[]
%     \centering
%         \caption{Model Validation Error metrics in $\%$ for IBR in GFM configuration}
%     \begin{tabular}{c|c|c|c|c|c|c}
%     \hline
%     \hline
%         Response & IBR- & IBR- & IBR- & IBR- & IBR- & all \\
%         error & 1 & 2& 3 & 4 & 5 & \\
%         \hline
%         $e_{v_{\mu}}$& 1 & 2& 3 & 4 & 5 & all\\
%         \hline
%         $e_{v_{\sigma}}$& 1 & 2& 3 & 4 & 5 & all\\
%         \hline
%         $e_{f_{\mu}}$& 1 & 2& 3 & 4 & 5 & all\\
%         \hline
%         $e_{f_{\sigma}}$& 1 & 2& 3 & 4 & 5 & all\\
%         \hline
%     \end{tabular}
%     \label{tab:err_GFM_table}
% \end{table}

\begin{table}
\renewcommand{\arraystretch}{1.2}
\caption{Power electronics-based load models representing house appliances }
\label{tab:load_models}
\centering
\begin{tabular}{l l}
\hline
 Load model & House appliances \\
\hline
Rectifier + Buck DC-DC converter & Desktop, home entertainment\\
Rectifier + Flyback DC--DC converter & Laptop charger \\
VFD + Induction motor & HVAC, washer, dryer \\
Boost converter + inverter & PV system, EV charger\\
\hline
\end{tabular}
\vspace{-3mm}
\end{table}

\begin{table}[]
\caption{Voltage variations to emulate different fault conditions in the distribution network}
\label{tab:vol-variations}
\begin{tabular}{|c|c|c|c|c|}
\hline
Case & \begin{tabular}[c]{@{}c@{}}Voltage before\\ dip (p.u.)\end{tabular} & \begin{tabular}[c]{@{}c@{}}Voltage during\\ dip (p.u.)\end{tabular} & \begin{tabular}[c]{@{}c@{}}Cycles of \\voltage dip\end{tabular} & \begin{tabular}[c]{@{}c@{}}Voltage after \\dip (p.u.)\end{tabular} \\ \hline
1 & 1 & 0.6 & 6 & 1 \\ \hline
2 & 1 & 0.5 & 10 & 1 \\ \hline
3 & 1 & 0.5 & 10 & 0.8 \\ \hline
4 & 1 & 0.5 & 6 & 0.8 \\ \hline
\end{tabular}
\end{table}

\begin{figure}
    \centering    \includegraphics[clip,width=80mm,scale=0.8]{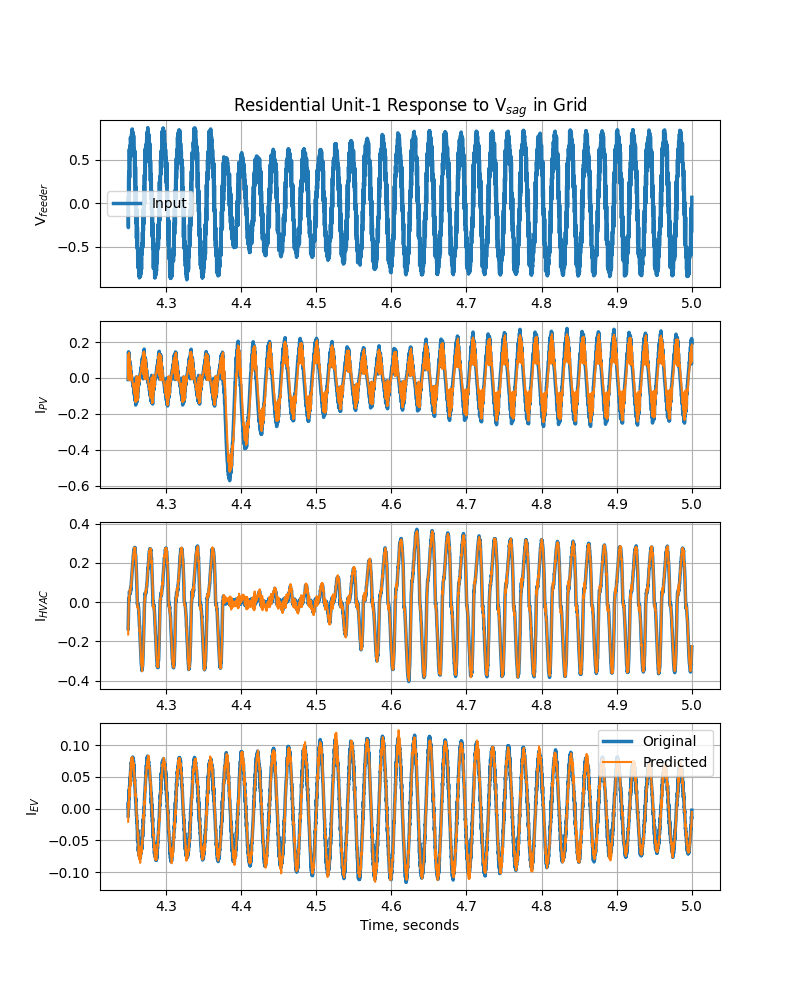}
    \caption{EMT domain modeling test results in the presence of past measurements are shown here. Given the system voltage, the loading of a residential unit is estimated. The predicted response and actual responses for HVAC, PV, EV currents are plotted.}
    \vspace{-20pt}
    \label{fig:EMT_signal_true_pred_case1}
\end{figure}

\begin{figure}
    \centering    \includegraphics[clip,width=80mm,scale=0.8]{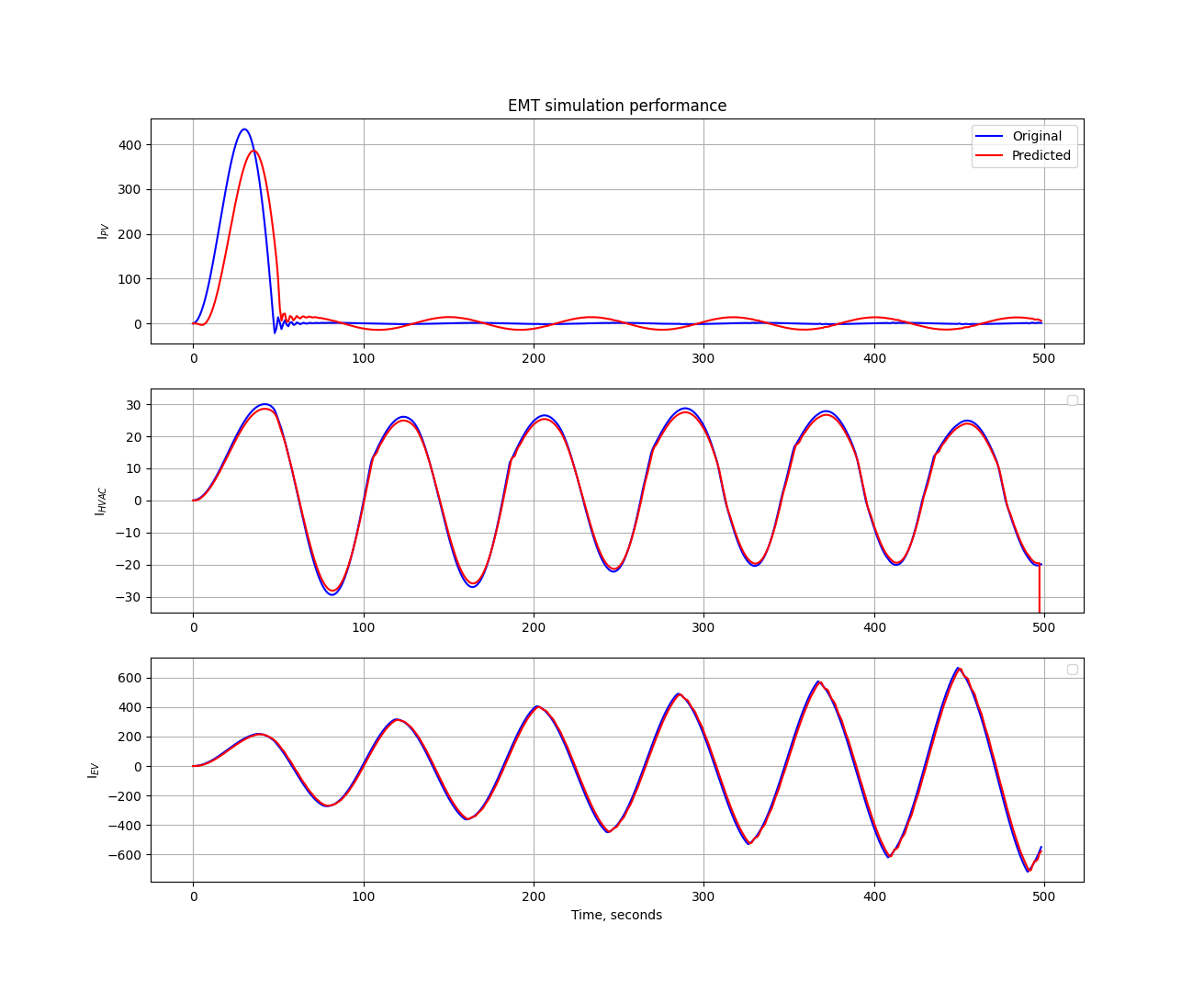}
    \caption{EMT domain modeling test results in the absence of past measurements are shown here. Given the system voltage, the loading of a residential unit is estimated. The predicted response and actual responses for HVAC, PV, EV currents are plotted.}
    \label{fig:EMT_signal_true_pred_case2}
\end{figure}

% \subsubsection{Test system description for EMT model}

\subsection{EMT domain modeling of IBR}

\subsubsection{Test system description}
For EMT domain modeling of IBR, a test system was created which consists of a distribution feeder with a set of residential customer loads and corresponding PV generation units. Residential customers are supplied through a single split-phase connection in the test system. Two houses having similar load composition have been connected across a 7.2kV/240V split-phase distribution transformer. The loads' compromise of power electronic load combinations and have been described in Table \ref{tab:load_models} \cite{mitra2023analyzing}. Detailed power electronic models for each of the described loads were designed in PSCAD/EMTdc. Voltage and current measurements were recorded at several locations as shown in Fig. \ref{fig:EMT}. Variations in voltage to replicate fault conditions upstream from the feeder were created using a custom-designed voltage dip as discussed in \cite{liu2018impact}. The different voltage variations are summarized in Table \ref{tab:vol-variations}. Data on the current and voltage phase point on wave (POW) data at each house, along with the current POW drawn or injected by each load at a sampling frequency of 5kHz was recorded. Additionally, the voltage at both the primary and secondary sides of the split-phase transformer was also recorded. the major load currents drawn from the household loads (HVAC), photovoltaic (PV) IBR units, and electric vehicle (EV) charging were recorded. 

The recordings were obtained under feeder voltage variations and voltage sag in a set of 4 scenarios. Each scenario data is then preprocessed with normalization and scaling to prepare the dataset for ARMAX model training and testing. One scenario data was used for training and rest 3 were used for testing. We conducted two different types of testing in this work for EMT domain calibration and validation. (1) In the presence of local measurements, the measurement data is continuously fed into the estimated model as output feedback. (2) In the absence of local measurements, the estimated output of the model is continuously fed into the UDM as output feedback. 

%The first one shows the continual validation case, while the second one conveys if the incoming measurement data is lost, how to still predict the response of IBR. 

\subsubsection{EMT domain modeling results}
For continual validation case, the past measurements are assumed to be available for doing this experiment and estimating validation error. The ARMAX model response is continuously compared with that of the original EMT simulation response at PV, EV, and HVAC locations. 
The results are presented in Figure. \ref{fig:EMT_signal_true_pred_case1}. The response shows a voltage sag event occurring at the distribution feeder bus and all the currents from the residential unit respond to this event. As can be seen, the estimated response of the ARMAX model accurately tracks the currents drawn by these units similar to that of its EMT simulation. 

Similarly, estimation results in the absence of incoming measurements are also presented in Figure \ref{fig:EMT_signal_true_pred_case2} which shows absence of measurements from time step 300, the estimated model was used to continuously predict output currents based on its past output response and thus is accurate up to  500 time steps after which the estimation error increases. This also suggests the re-estimation of ARMAX model needed when the error increases to maintain accuracy of the model and making online model updates.
The average RMSE across 4 different scenarios were calculated and presented in Table \ref{tab:err_EMT_table} which shows the average error in time series response estimation both in case of presence and absence of measurements to be within 0.5-1.5\% of the original current values.

% % The following p

% \begin{figure}
%     \centering   
%     \includegraphics[clip,width=80mm,scale=0.6]{Figures/Ind2_IBR_case_4_inv_0_v0.png}
%     \caption{Caption}
%     \label{fig:enter-label}
% \end{figure}

% \begin{figure}
%     \centering    
%     \includegraphics[clip,width=80mm,scale=0.6]{Figures/err_Ind_IBR_case_4_inv_0_v0.PNG}
%     \caption{Caption}
%     \label{fig:enter-label}
% \end{figure}

\begin{table}[]
    \centering
        \caption{Model Validation Error metrics in $\%$ for IBR in EMT simulation-based modeling}
    \begin{tabular}{|c|c|c|c|}
    % \hline
    \hline
        Response & PV & HVAC & EV  \\
        Error & current & current & current \\
         % &   &   &   &   \\
        \hline
        % Case & 0.01 & 0.01 & 0.005  \\
        Average & 1.5 & 1.5 & 0.5  \\
        \hline
    \end{tabular}
    \label{tab:err_EMT_table}
\end{table}

\section{Conclusion}
This paper presents a framework for UDM calibration and validation of multiple IBR models both in GFM and GFL configuration and both in phasor and EMT domain. Proposed data driven modeling of the IBR using ARMAX model is shown to accurately match the response of a generic IBR model \cite{esig_generic} in both phasor domain and EMT domain simulation time steps. Proposed UDMs were validated for a large set of dynamic event simulations and the results in terms of RMSE across multiple reponses were presented. Continual validation result shows good accuracy of the proposed ARMAX models both in the presence and absence of incoming measurements. This implies the proposed model can be used to predict response of not only IBRs but also residential loads/generation currents such as EV, PV, and HVAC loads with great precision.  This type of data-driven modeling can be highly beneficial for optimizing and managing these resources. The proposed model is expected to help in resource optimization, solar forecasting, control and digital twin applications where response of IBRs need to be predicted. Future direction in this research would include testing the proposed modeling framework for field measurements obtained from IBR locations. 

\section*{Acknowledgment}
This work was supported by the Sensors and Data Analytics Program of the U.S. Department of Energy Office of Electricity, under Contract No. DE-AC05-76RL01830. The authors gratefully acknowledge Dr. Rohit Jinsiwale and Dr. Aaqib Peerzada of the Pacific Northwest National Laboratory for their helpful suggestions and feedback.

\bibliographystyle{IEEEtran}

\bibliography{references_bib}

\begin{thebibliography}{10}
\providecommand{\url}[1]{#1}
\csname url@samestyle\endcsname
\providecommand{\newblock}{\relax}
\providecommand{\bibinfo}[2]{#2}
\providecommand{\BIBentrySTDinterwordspacing}{\spaceskip=0pt\relax}
\providecommand{\BIBentryALTinterwordstretchfactor}{4}
\providecommand{\BIBentryALTinterwordspacing}{\spaceskip=\fontdimen2\font plus
\BIBentryALTinterwordstretchfactor\fontdimen3\font minus
  \fontdimen4\font\relax}
\providecommand{\BIBforeignlanguage}[2]{{%
\expandafter\ifx\csname l@#1\endcsname\relax
\typeout{** WARNING: IEEEtran.bst: No hyphenation pattern has been}%
\typeout{** loaded for the language `#1'. Using the pattern for}%
\typeout{** the default language instead.}%
\else
\language=\csname l@#1\endcsname
\fi
#2}}
\providecommand{\BIBdecl}{\relax}
\BIBdecl

\bibitem{nerc2021odessa}
{Joint NERC and Texas RE Staff Report}, ``{Odessa Disturbance: Texas Events:
  May 9, 2021 and June 26, 2021},'' {North American Reliability Coordination
  (NERC)}, Tech. Rep., Sep 2021.

\bibitem{nerc2022odessa}
------, ``{2022 Odessa Disturbance: Texas Event: June 4, 2022},'' {North
  American Reliability Coordination (NERC)}, Tech. Rep., Dec 2022.

\bibitem{nerc2020BPS}
{NERC Staff Report}, ``{BPS}-connected inverter-based resource modeling and
  studies, technical report, may 2020,'' 2020.

\bibitem{esig_generic}
W.~{REMTF}, ``Generic models ({WPPs}),''
  \url{https://www.esig.energy/wiki-main-page/generic-models-wpps/}, accessed:
  2024-06-25.

\bibitem{km_gen_model}
K.~Mahapatra and H.~Wang, ``Generator dynamic model calibration using multiple
  disturbance events,'' in \emph{2020 IEEE Power \& Energy Society Innovative
  Smart Grid Technologies Conference (ISGT)}, 2020, pp. 1--5.

\bibitem{ju2020indices}
W.~Ju, N.~Nayak, C.~Vikram, H.~Silva-Saravia, K.~Sun, and G.~Zu, ``Indices for
  automated identification of questionable generator models using
  synchrophasors,'' in \emph{2020 IEEE Power \& Energy Society General Meeting
  (PESGM)}.\hskip 1em plus 0.5em minus 0.4em\relax IEEE, 2020, pp. 1--5.

\bibitem{ferc_901}
``{Federal Energy Regulatory Commission} ({FERC}) 901 - {R}eliability standards
  to address inverter-based resources,''
  \url{www.ferc.gov/media/e-1-rm22-12-000}, accessed: 2024-06-25.

\bibitem{kim2020machine}
S.~H. Kim and F.~Boukouvala, ``Machine learning-based surrogate modeling for
  data-driven optimization: a comparison of subset selection for regression
  techniques,'' \emph{Optimization Letters}, vol.~14, no.~4, pp. 989--1010,
  2020.

\bibitem{dig_twin}
F.~Tao, B.~Xiao, Q.~Qi, J.~Cheng, and P.~Ji, ``Digital twin modeling,''
  \emph{Journal of Manufacturing Systems}, vol.~64, pp. 372--389, 2022.

\bibitem{fan2020time}
L.~Fan and Z.~Miao, ``Time-domain measurement-based $ dq $-frame admittance
  model identification for inverter-based resources,'' \emph{IEEE Transactions
  on Power Systems}, vol.~36, no.~3, pp. 2211--2221, 2020.

\bibitem{yang1995identification}
H.-T. Yang, C.-M. Huang, and C.-L. Huang, ``Identification of armax model for
  short term load forecasting: an evolutionary programming approach,'' in
  \emph{Proceedings of Power Industry Computer Applications Conference}.\hskip
  1em plus 0.5em minus 0.4em\relax IEEE, 1995, pp. 325--330.

\bibitem{wecc_arma}
W.~I. M.~R. Group \emph{et~al.}, ``{Modes of Inter-Area Power Oscillations in
  the Western Interconnection},'' 2021.

\bibitem{singh2013report}
A.~K. Singh and B.~C. Pal, ``Report on the 68-bus, 16-machine, 5-area system,''
  2013.

\bibitem{lasseter2010certs}
R.~H. Lasseter, J.~H. Eto, B.~Schenkman, J.~Stevens, H.~Vollkommer, D.~Klapp,
  E.~Linton, H.~Hurtado, and J.~Roy, ``Certs microgrid laboratory test bed,''
  \emph{IEEE Transactions on Power Delivery}, vol.~26, no.~1, pp. 325--332,
  2010.

\bibitem{du2020modeling}
W.~Du, F.~K. Tuffner, K.~P. Schneider, R.~H. Lasseter, J.~Xie, Z.~Chen, and
  B.~Bhattarai, ``Modeling of grid-forming and grid-following inverters for
  dynamic simulation of large-scale distribution systems,'' \emph{IEEE
  Transactions on Power Delivery}, vol.~36, no.~4, pp. 2035--2045, 2020.

\bibitem{kwon2023risk}
K.-b. Kwon, S.~Mukherjee, T.~L. Vu, and H.~Zhu, ``Risk-constrained
  reinforcement learning for inverter-dominated power system controls,''
  \emph{IEEE Control Systems Letters}, 2023.

\bibitem{sippy_paper}
G.~Armenise, M.~Vaccari, R.~B. Di~Capaci, and G.~Pannocchia, ``An open-source
  system identification package for multivariable processes,'' in \emph{2018
  UKACC 12th International Conference on Control (CONTROL)}, Sep. 2018, pp.
  152--157.

\bibitem{mitra2023analyzing}
B.~Mitra, A.~Singhal, S.~Kundu, and J.~P. Ogle, ``Analyzing distribution
  transformer degradation with increased power electronic loads,'' in
  \emph{2023 IEEE Power \& Energy Society Innovative Smart Grid Technologies
  Conference (ISGT)}.\hskip 1em plus 0.5em minus 0.4em\relax IEEE, 2023, pp.
  1--5.

\bibitem{liu2018impact}
Y.~Liu, Y.~Zhang, Q.~Huang, S.~Kundu, Y.~Tang, D.~James, P.~Etingov, B.~Mitra,
  and D.~P. Chassin, ``Impact of building-level motor protection on power
  system transient behaviors,'' in \emph{2018 IEEE Power \& Energy Society
  General Meeting (PESGM)}.\hskip 1em plus 0.5em minus 0.4em\relax IEEE, 2018,
  pp. 1--5.

\end{thebibliography}

\end{document}